\documentstyle[psfig,aaspp4]{article}

\newcommand{\mdust}{M_{\rm dust}}
\newcommand{\tdust}{T_{\rm dust}}
\newcommand{\fsix}{f_{60}}
\newcommand{\fhun}{f_{100}}
\newcommand{\mhi}{M_{\rm HI}}
\newcommand{\mhiopt}{M_{\rm HI,opt}}
\newcommand{\fhi}{S_{HI}}

\newcommand{\msun}{M_{\odot}}

\newcommand{\hi}{H\,{\sc i}\ }
\newcommand{\hii}{H\,{\sc ii}\ }
\newcommand{\halpha}{{\rm H_\alpha}}
\def\etal{{et al.~\/}}

\def\ie{{i.e.~\/}}

\def\ltsima{$\; \buildrel < \over \sim \;$}
\def\simlt{\lower.5ex\hbox{\ltsima}}
\def\gtsima{$\; \buildrel > \over \sim \;$}
\def\simgt{\lower.5ex\hbox{\gtsima}}

\lefthead{Lisenfeld \&  Ferrara}
\righthead{Dust and Metals in Dwarf Galaxies}

\begin{document}
\title{Dust-to-gas Ratio and Metal Abundance in Dwarf Galaxies}
\author{U. Lisenfeld\altaffilmark{1} \& A. Ferrara\altaffilmark{2}}
\affil{
Osservatorio Astrofisico di Arcetri\\
Largo E. Fermi 5 \\ 50125 Firenze, Italy}
\altaffiltext{1}{Present address:
Universidad de Granada, Departamento de F\'\i sica
Te\'orica y del Cosmos, 18002 Granada, Spain, E-mail: ute@ugr.es}
\altaffiltext{2}{E--mail: ferrara@arcetri.astro.it}

\begin{abstract}
We have compared the metallicity (represented by
oxygen abundance), $X_o$, and the dust-to-gas ratio, ${\cal D}$, 
in a sample of dwarf  galaxies.  For dwarf irregulars (dIrrs) 
we find a good correlation between the two quantities, with a power-law index 
${0.52\pm 0.25}$. Blue Compact Dwarf
(BCD) galaxies do not show such a good correlation; in addition
both the dust-to-gas ratio and the metallicity tend
to be higher than for dIrrs. We have then developed a simple but physical
analytical model for the above relation. Comparing the model results with 
the data, we conclude that: 
(i) for low values of ${\cal D}$, the ${\cal D}-X_o$
relation is quasi-linear, whereas for higher values the curve
strongly deviates from the linear behavior,
implying that the commonly used power-law approximation is very
poor; 
(ii) the deviation from the linear behavior depends critically on the
parameter $\chi$, the ``differential'' mass outflow rate from
the galaxy in units of the star formation rate, $\psi$; 
(iii) the {\it shape} of
the ${\cal D}-X_o$ curve does not depend on  $\psi$, but only on  $\chi$;
however, the specific location of a given galaxy on the curve does depend on $\psi$;
(iv) the BCD metallicity segregation is due to a higher $\psi$, 
together with a significant differential mass outflow.
Thus, the lack of correlation can be produced by largely different 
star formation rates and values of $\chi$ in
these objects.
\end{abstract}


\section{Introduction}

Interstellar dust consists mainly of heavy elements. Therefore
a relation between the dust-to-gas ratio and the metallicity
of a galaxy can be expected. The simplest argument to derive
this dependence comes from Franco \& Cox (1986): the authors       
suggest that the dust-to-gas ratio, ${\cal D}$, is simply $\propto Z$, the 
metallicity of the gas.
However, this assumption might be too simplistic.
Furthermore, the dust-to-gas ratio might be expected to depend on
the star formation rate. 

Dwek \& Scalo (1980) have shown that dust injection from supernovae dominates
other sources, if indeed grains can form and survive in the ejecta.
This has become clear after the SN1987A event, in which dust has been
unambiguously detected (Moseley \etal 1989, Kosaza, Hasegawa \&
Nomoto 1989). 
Indeed, the bulk of the refractory elements (characterized by higher melting
temperatures, such as Si, Mg, Fe, Ca, Ti, Al etc.)
is injected into the ISM by supernovae (McKee 1989).
Dust can be destroyed by interstellar shocks: small grains are destroyed
by thermal sputtering in fast non-radiative shocks; large grains are
destroyed by grain-grain collisions and eroded by non-thermal sputtering
in radiative shocks (Jones \etal 1994, Borkowski \& Dwek 1995). In addition, dust grains
might be expelled from the galaxy either by radiation pressure
(Ferrara \etal 1991) or convected away by a fountain-like 
flow (Norman \& Ikeuchi 1989). From these arguments, it follows that
the dust content of a galaxy is clearly governed by supernovae, and
therefore the equilibrium established between the 
formation and destruction processes could be influenced by the star 
formation history of the galaxy.

The aim of this paper is to investigate these processes.
As a first step we want to ascertain empirically the existence of a relation 
between the dust-to-gas ratio, metallicity and the star formation activity.
The results are used to constrain a model for the dust production
taking into account the processes described above.
We have selected dwarf galaxies as objects for this study. The reason
is twofold: (i) the metallicities
observed in these objects span a large range of values;
(ii) due to their
small size and simple structure, radial gradients in the dust-to-gas
ratio and metallicity can be neglected. Both features make
them suitable objects for a comparison of the dust-to-gas ratio and 
metallicity.
The knowledge of the dust content in dwarf galaxies is also relevant
to cosmological studies as dwarfs represent the most numerous
population of galaxies.
Thus, their dust content could have important implications, e.g.
for the obscuration of QSOs. 

Detailed studies of the extinction properties of the Magellanic
Clouds have revealed a clear relation  
between the dust-to-gas ratio and the metal abundance:
In the Large Magellanic Cloud (LMC) both the metallicity 
and the dust-to-gas ratio are 
roughly a factor 4 lower than in the Galaxy
(Koornneef 1982; Fitzpatrick 1986) 
and in the Small Magellanic Cloud (SMC), where the 
metallicity is a factor of about 10 lower than in the Galaxy,
also the dust-to-gas ratio is a factor of 8 - 11 lower (Bouchet \etal 1985). 
So far only few studies have investigated the
relation between the dust-to-gas ratio and the metallicity 
for a 
large number of galaxies. 
Issa, MacLaren \& Wolfendale (1990) have compared the dust-to-gas ratio and the
metallicities in the Galaxy and 6 nearby galaxies and found
evidence for a correlation, with the two quantities decreasing at
roughly the same rate with increasing galactocentric radius. 
When taking both quantities at a fixed radius they also obtained
a good correlation, following 
$Z \sim {\cal D}^{0.85\pm 0.1}$.             
Schmidt \& Boller (1993) have compared the dust-to-gas ratio
and metallicity of  23 nearby dwarf galaxies and found that the
oxygen abundance is $\sim {\cal D}^{0.63\pm 0.25}$, hence in good 
agreement with Issa \etal (1990).

In the present work we have compiled
data for dwarf irregular (dIrr) galaxies and blue compact
dwarf (BCD) galaxies from the literature and studied the
relation between the dust and metal abundance and the star 
formation rate.  For the dIrr sample we confirm the 
existence of a correlation which is clearly non-linear whereas 
for the BCD sample no significant correlation could be found.
These results  allow a theoretical insight on the physics
that governs the metal and dust production and on global galactic properties
such as the star formation and mass loss history.

\section{The samples and the data}

We divide the dwarf galaxies in our study into two subsamples: one sample 
contains dIrr galaxies 
and the other sample consists of BCDs. In the dIrr sample we
include all galaxies of morphological type S(B)m or I(B)m 
without applying a, in any case arbitrary, luminosity cutoff.
The reason for distinguishing between dIrrs and BCDs
is their different recent star formation history: 
BCDs possess spectra  very similar to \hii regions, and they
are believed to be undergoing a starburst.
This variable, intense star formation rate in the recent past could affect
the relation between the dust and metal abundance, as the time scale for
a starburst ($\approx 10^7-10^8$ years)
is comparable to the time scales relevant for the dust
and metal production which is given by the life time of massive stars.
Most of the dIrrs, on the other hand, are not 
in such an extremely intense star forming phase. 

In Tab. 1 and 2 the data for the samples are  listed.
The sample of dIrrs consists
of 28 galaxies. The sample size was mainly constrained by
the availability of metallicity data.
The oxygen abundances, which were chosen as a tracer for the metallicity,
were taken from a compilation  
of Skillman, Kennicutt \& Hodge (1989), 
Hunter, Gallagher \& Rautenkranz (1982) (corrected according to
McGaugh 1991),
Campbell, Terlevich \& Melnick (1986)  and
Schmidt \& Boller (1993). The latter authors did a 
similar study to the present one; their sample
contains 23 dIrr galaxies and is a subsample of ours.
The rest of the data (far-infrared (FIR) flux, distance, blue luminosity, 
\hi mass, optical diameter) were taken from the 
catalogue of Melisse \& Israel (1994a)
except for NGC 4236 which was not included in their sample.
For this galaxy we took the blue magnitude from de Vaucouleurs et
al. (1991) (Third Reference Catalogue, RC3) and the \hi flux from 
Huchtmeier \& Richter (1989). 

The BCD sample contains 16 galaxies. 
We have restricted the sample to small galaxies (optical diameter less
than 10 kpc) in order to obtain a sample of galaxies comparable
in size to the dIrrs.
The metallicities and distances were taken from 
Kunth \& S\`evre (1986), Masegosa, Moles \& Campos-Aguilar (1994) 
and Terlevich \etal (1992).
The \hi fluxes are from the Huchtmeier \& Richter (1989), Huchtmeier,
Hopp \& Kuhn (1997), Stavely-Smith, Davies \& Kinman (1992) and from
RC3. The FIR
fluxes were taken from the IRAS Faint Source Catalogue (1990).

The \hi mass is derived from the \hi flux, $S_{HI}$ as:
\begin{equation}
\mhi  =2.36\,10^5 \left({\fhi \over \rm Jy\,km\,s^{-1}} \right)
\left({D\over \rm Mpc}\right)^2 \msun.
\end{equation}

\subsection{Calculation of the dust mass}

The dust masses were calculated from the FIR emission using a single temperature, black-body model:
\begin{equation}
\mdust={\fhun D^2 \over  B(\nu,\tdust) Q_\nu}
\end{equation}
where $B(\nu,\tdust)$ is the Planck emission function, $D$ the distance to 
the galaxy, $\fhun$ the IRAS flux at 100$\mu$m and $\tdust$ the
dust temperature.
The infrared  dust absorption coefficient, $Q_\nu$, is assumed to be the
same for all galaxies and was taken from Hildebrand (1983):
\begin{equation}
Q_\nu=0.1 \left({250\over \lambda}\right)^{\beta} {\rm g/cm}^2;
\end{equation}
$\tdust$ is calculated from the ratio of the
IRAS fluxes at 60 and 100 $\mu$m.
Together this yields:
\begin{equation}
\mdust =12\,\left({\fhun\over \rm Jy}\right)
   \left({ D\over \rm Mpc}\right)^2 \left({250\over 100}\right)^{-\beta}
 \left\{ \left[ {\fhun\over\fsix} \left({100\over 60}\right)^{\beta+3}
\right]^{1.5} -1 \right\}    \msun
\end{equation}
%
%
We assumed $\beta=2$. The choice of $\beta$ has
only a minor influence on the dust mass.

\subsection{Uncertainties in the dust mass calculation}

The dust mass inferred from the FIR emission as described above is affected
essentially by two main uncertainties. 

First, the value of the dust mass calculated as
described above depends sensitively on the dust
temperature derived from the 60-to-100$\mu$m flux ratio. 
We assume hereby that both the 60 and the 100$\mu$m flux is dominated
by the emission of big grains, that contain most of
the dust mass, and are in thermal equilibrium
with the radiation field.  
If very small grains (VSGs), whose 
temperature is fluctuating, 
contribute
significantly to the 60$\mu$m emission this method produces wrong
results. 
An analysis of the Galactic diffuse, 'cirrus`, emission and extinction
curve in the solar neighborhood (D\'esert, Boulanger \& Puget 1990)
has shown that the emission of 
VSGs could indeed  be responsible for 60\% of the 60$\mu$m flux, 
but only for 15\%
of the flux at 100$\mu$m. Since  the contribution
of VSGs to the dust mass is negligible, 
this results in an error in the dust mass of a factor 3. 
A similar result was obtained by Sodrowski \etal (1987) who
analyzed the FIR emission in our Galaxy and found that
the dust temperature derived from the 60/100$\mu$m ratio stays almost
constant with Galactic radius in spite of the decreasing radiation field.
A possible interpretation is that the 60$\mu$m emission is 
to a large extent produced by  VSGs.
The apparent dust-to-gas mass ratio derived from the FIR emission with
this constant temperature profile decreased by a factor of 3 outwardly.
Assuming that the dust-gas ratio is the
same everywhere in the Galaxy, the  implied error of the dust mass
in this analysis is also a factor of 3. 

\begin{figure}
\psfig{file=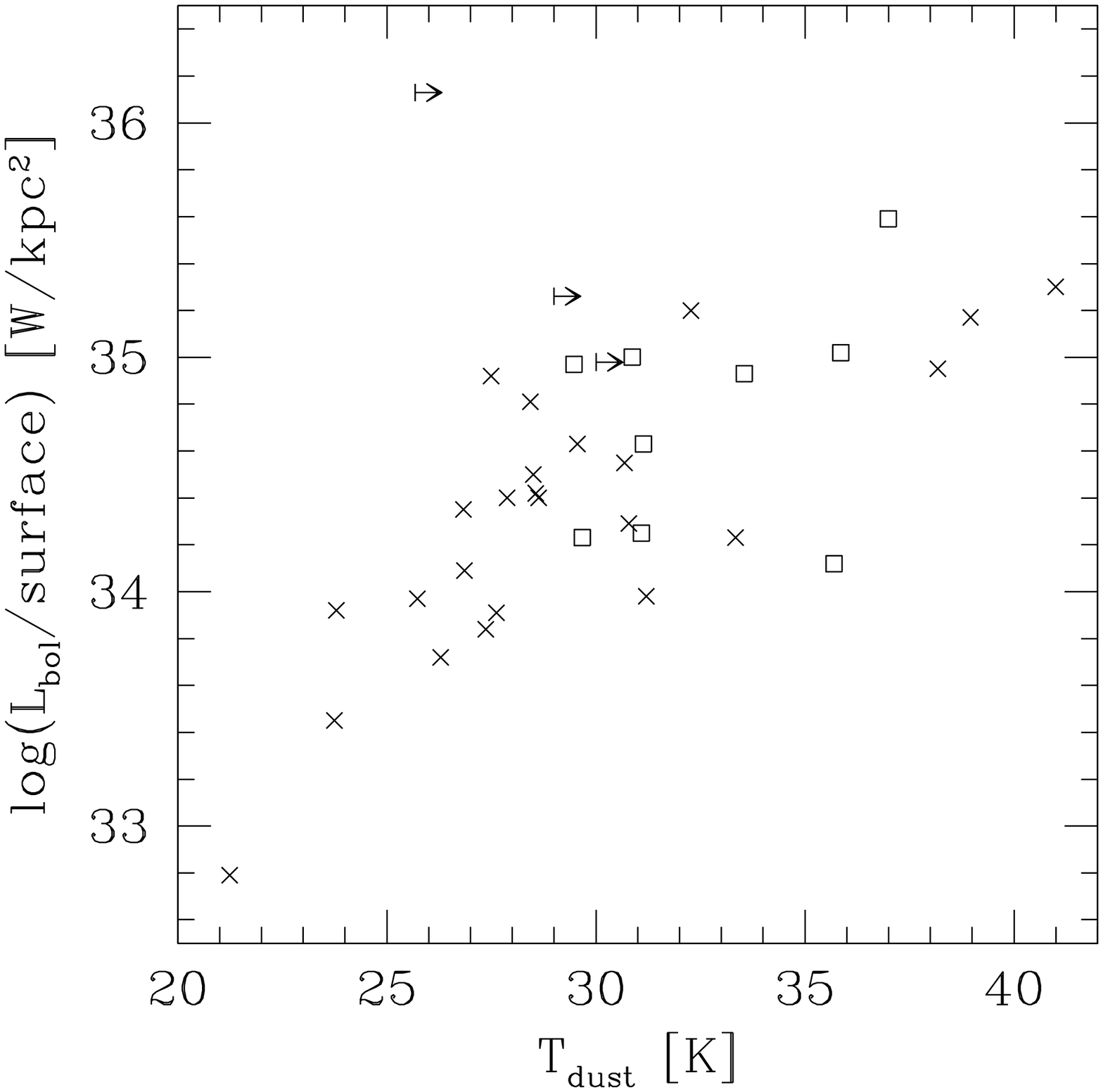,width=8.5cm,clip=,angle=0}
\caption{\label{bol-temp} Bolometric surface brightness vs.
dust temperture for both samples.  The open squares denote 
BCDs with FIR detections,
the arrows BCDs with upper limits in the 100 $\mu$m flux, and the crosses
dIrr galaxies with FIR detections.}
\end{figure}

However, the contribution of VSGs to the 60$\mu$m flux is expected to be
only significant in a low radiation field in which the
big grains have a low temperature and contribute little to the 60$\mu$m
emission (D\'esert \etal 1990). This is not the case for dwarf galaxies which
show higher
60/100$\mu$m and lower 12/25$\mu$m ratios than spiral galaxies
(Xu \& de Zotti 1991, Melisse \& Israel 1994b) indicating
a lack of 'cirrus' emission.
Therefore we can expect that
the error due to VSGs is smaller than estimated for the Galactic
cirrus emission. We have carried out the
following qualitative test for this assumption: 
If the dust temperature derived from the 60 and 100$\mu$m
fluxes is a good measure
for the real temperature of the big grains, it should correlate
with the radiation field.  In Fig. \ref{bol-temp} the dust temperature and 
the surface brightness of the bolometric luminosity for
both samples are plotted. The bolometric luminosity was calculated
as described below and the galaxy surface 
area is obtained from the optical diameter.
A good correlation 
between the dust temperature and the bolometric surface brightness
is visible. The correlation coefficient (only detections) is
$r=0.75$. This supports our assumption that
the dust temperature represents the temperature of big grains 
in thermal equilibrium.

Next, the dust-to-gas ratio derived from the FIR 
emission is generally significantly, up to a factor of 10, lower than the
value found from the analysis of the extinction 
in the Solar Neighborhood. This has been found 
both for simple blackbody models like the 
present one (Young \etal 1989; Devereux
\& Young 1990) as well as for more detailed dust models (Rowan-Robinson
1992).
The presence of cold dust, emitting mainly beyond 100 $\mu$m, might be 
responsible for the discrepancy. On the basis of the 
available data it is not possible to decide about this
point, observations at longer wavelengths would be necessary.
Because of this uncertainty 
we have carried out   
a further test to check whether our dust masses are reasonable:
We have compared them
with an alternative way to estimate the dust mass, through 
an energy balance between the fraction of the radiation 
absorbed by the dust and escaping
from a galaxy (Xu \& Buat 1995; Xu \& Helou 1996). 
From the fraction of 
nonionizing radiation absorbed by the dust 
we can estimate the dust opacity and from that, 
assuming that the opacity produced by a certain
dust mass is the same as in the Galaxy, the dust
mass per area can be calculated. 
We have calculated the fraction of radiation absorbed by dust
following the model of Xu \& Buat (1995) who estimate the bolometric
luminosity from the ultraviolet ($f_{UV}$), 
blue ($f_B$) and FIR fluxes. For our sample, the ultraviolet (UV)
fluxes were available for only 9 galaxies (taken from Fanelli \etal 1997; 
Hill \etal 1994; Deharveng \etal 1994;
Donas \etal 1987; Donas \& Deharveng 1984). For these we 
derived $f_B/f_{2000 \AA}=1.0$  which we use for the whole sample. 

The ratio of the FIR-to-bolometric luminosity gives
the fraction of radiation absorbed by the dust. If our dust
mass estimate is correct, this value should be related to
the dust mass per surface area. Indeed, we found a good correlation between
these two quantities ($r=0.87$).
In order to quantify this further we have to make a distinction between
dust emission from \hii regions and diffuse dust emission. 
The dust in \hii regions is mainly heated by ionizing photons which
are almost completely absorbed by the dust locally, practically independent
of the dust amount. The diffuse dust, on the other hand,  
is mainly heated by nonionizing
photons and the amount of radiation absorbed is directly related to the 
dust opacity.
Thus, in order to calculate the (diffuse) dust opacity we have to subtract the
dust heating by ionizing photons. 
We estimate the ionizing UV radiation from the $\halpha$ emission.
The total $\halpha$ fluxes were available for a large part of the dIrr sample
(18  galaxies, $\halpha$ fluxes taken from Hunter, Hawley \& Gallagher 1993,
Hunter \etal 1989, Kennicutt \& Kent 1983). 
Since the dust opacity is generally low in dwarf galaxies, 
the extinction of the $\halpha$ line is likely to be small and we neglected it.
The total dust emission 
originating from ionizing radiation can be estimated as
$f_{\rm dust}^{\rm lyc}=27.12 \times f_{\halpha}$
(Xu \& Buat 1995). 
The average fraction of dust emission due to heating by ionizing photons
is $f_{\rm dust}^{\rm lyc}/f_{\rm dust}=0.45 \pm0.35$,
where  $f_{\rm dust}$ is the 
total dust emission, extrapolated from the FIR emission according to 
Xu \& Buat (1995).
We use this average value for the galaxies for which no $\halpha$
data is available. 

The diffuse dust opacity is estimated using a 
simplified radiation transfer model
with a slab geometry, assuming that dust and stars are homogeneously
mixed. 
In view of the uncertainty of the real diffuse dust distribution
in the individual galaxies
this simplification  seems to be justified.
A different geometry, e.g. a 'sandwich distribution` or  
a clumpy dust distribution would lead to a different value for the
dust opacity and the dust mass. A crucial assumption in the
present test is therefore that  there is no {\it systematic} change in
the diffuse dust morphology, i.e. that there is no correlation between
the geometry (e.g. clumpiness) and the 
absorbed  fraction of radiation,  dust-to-gas ratio or metallicity.
We use an approximate formula 
(Xu \& de Zotti 1989) for the dust absorption probability, 
further simplified by dividing the nonionizing radiation into
two (optical and UV) bands for which we take the
extinction properties at 4300 \AA \ and 2000 \AA, respectively
(see Lisenfeld, V\"olk \& Xu 1996).

\begin{figure}
\psfig{file=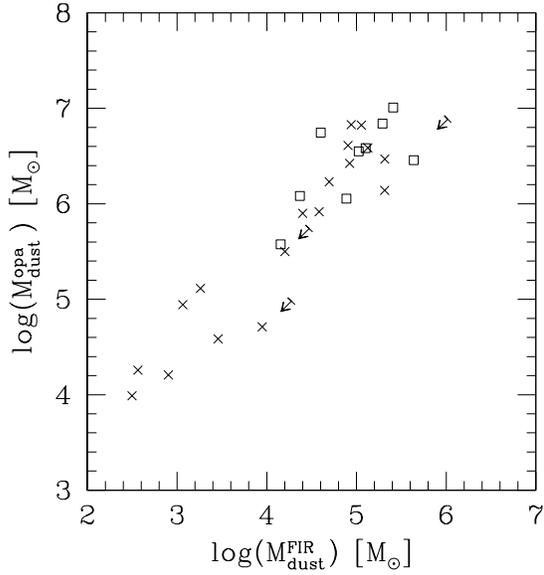,width=8.5cm,clip=,angle=0}
\caption{\label{tau-mass}  Dust masses calculated from the 60 and 100
$\mu$m fluxes ($M_{dust}^{FIR}$) and from the opacity ($M_{dust}^{opa}$).
The symbols are defined as in Fig. \ref{bol-temp}.}
\end{figure}

The opacities deduced (at 4300 \AA ), $\tau_B$, 
range between $0.01$ and $0.4$.
Assuming that the dust has the
same extinction properties as dust in our 
Galaxy (D\'esert \etal 1990),
we can estimate the dust mass as
\begin{equation}
M_{dust}^{opa}[\msun]=8.8\,10^4 \times\tau_B \times \left({A_{gal}
\over \rm kpc^2}\right)
\end{equation}
where $A_{gal}$ is the optical surface area of the galaxy.
In Fig. \ref{tau-mass} the dust mass estimated in this way and the
dust mass calculated from the FIR emission are compared.
The most striking result is the very good ($r=0.96$) and   
linear (slope = $1.0 \pm 0.05$) correlation between the
two quantities. The dispersion around the regression line is 0.34 dex,
corresponding to a factor of 2.2. 
The dust mass derived from the FIR emission is on average a factor of 29
lower than the mass derived from the opacity.
However, since the main aim of the present work is the comparison
of the dust-to-gas ratio with the
metallicity of galaxies, the absolute value
of the dust mass is only of secondary importance;
the crucial point 
is that the {\it ratio} of the dust masses is correct.
The linear correlation
between the dust masses derived in two different ways (from the
FIR emission and from the extinction) corroborates that this
latter assumption is correct.

\section{Comparison of the dust-to-gas ratio and the metallicity}

\begin{figure}
\psfig{file=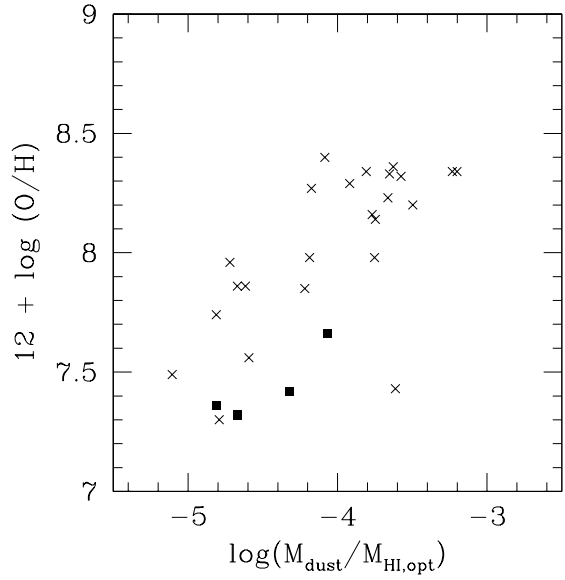,width=8.5cm,clip=,angle=0}
\caption{\label{dust-gas-dirr} 
Dust-to-gas-ratio and metallicity for the
dIrr sample. Filled squares
denote galaxies for which $\mdust$ is undetermined 
and only an indicative value is plotted (see text).
}
\end{figure}

\begin{figure}
\psfig{file=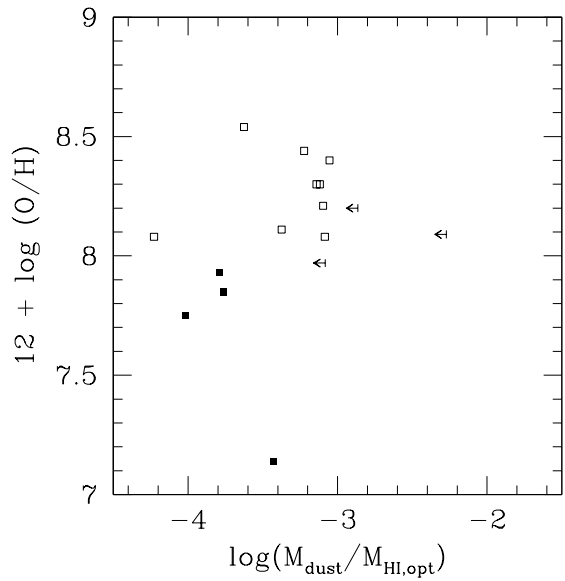,width=8.5cm,clip=,angle=0}
\caption{\label{dust-gas-bcd} Dust-to-gas-ratio and metallicity for the
 BCD sample. The arrows denote
upper limits in the 100 $\mu$m flux and filled squares
denote galaxies for which $\mdust$ is undetermined 
and only an indicative value is plotted (see text).}
\end{figure}

\begin{figure}
\psfig{file=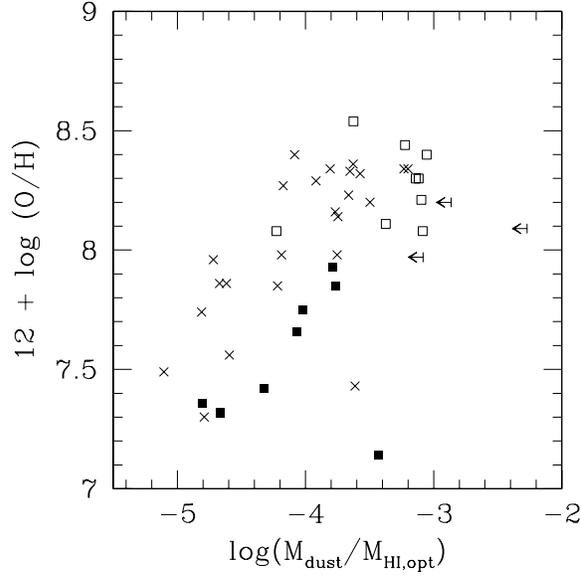,width=8.5cm,clip=,angle=0}
\caption{\label{dust-gas-tot} Dust-to-gas-ratio and metallicity for both
samples (dIrrs and BCDs) together. The symbols are defined as in Fig. \ref{bol-temp}
and  \ref{dust-gas-dirr}
}
\end{figure}

In Fig. \ref{dust-gas-dirr} - \ref{dust-gas-tot} 
the  dust-to-gas ratios and the metallicities
are plotted for the two samples.
For the galaxies with upper limits in both
IRAS fluxes, the
dust temperature is undefined and therefore, strictly
speaking, the dust mass as well. 
Since the dust temperature in the samples
shows a small dispersion (all values lie between 22 and
42 K)  we adopted  $T=30$ K for these 4 dIrr galaxies and
calculate the dust mass with this value in order to give 
at least an indicative value (filled squares in  in Fig. \ref{dust-gas-dirr}
- \ref{dust-gas-tot}). 
In the same way,  we adopt for 
the BCDs  that were not detected by IRAS an upper
limit of 1 Jy at 100 $\mu$m and a dust temperature of 32 K, corresponding
to the average value found for this  sample.

We only considered the  \hi gas mass in our analysis.
The extent of the \hi emission is generally much larger than
the optical diameter. Fouqu\'e (1983) 
has found  that on average for irregular galaxies
the \hi effective radius  (i.e. the radius within which half of the
\hi gas is contained), $D_{HI}$, and the optical isophotal radius, 
$D_{25}$, are nearly the same, $\log(D_{HI}/D_{25})=0.04 \pm 0.2$.
Even if there is dust associated 
with this extended \hi it is,
due to the absence of heating sources, very cold and therefore  
unlikely to contribute to the FIR emission.
Therefore, in order to get an accurate value for the dust-to-gas ratio
within the optical disk,  we only consider the
\hi mass within the optical disk, $\mhiopt$, by
dividing the total \hi mass by 2.

We could
not derive $\rm H_2$ masses for the whole sample because of the
lack of data.  
We expect the error to be small, at least for dwarf irregulars,
since the molecular gas content is generally much less than
the atomic gas content:
Whereas the molecular mass content of spiral galaxies is in general roughly 
comparable to the atomic gas content with a median value of
$M({\rm H_2})/M({\rm \hi})=0.5$ 
(Andreani, Casoli \& Gerin 1993; Devereux \& Young 1990),
in dwarf irregular this ratio is lower. 
Cohen \etal (1988) derived in the LMC  a 
molecular-to-atomic gas fraction 
of 30\%. Molecular hydrogen absorption measurements gave
an even lower value of 6 \% (Clayton \etal 1996).
For the SMC the molecular-to-atomic gas ratio is estimated to be 7\%
(Rubio \etal 1991).
%
For 7 galaxies of our sample
we could find CO data in the literature (Young \etal 1995,
Elfhag \etal 1996). 
The conversion factor from $\rm H_2$ to CO is extremely uncertain
for dwarf galaxies and very likely to be considerably higher than
the standard value used for Galactic molecular gas clouds.
The conversion factors found for the Magellanic Clouds are
6 times higher than for the Galaxy (LMC, Cohen \etal 1988)
and 20 times higher (SMC, Rubio \etal 1991).
Adopting an intermediate value of 
10 times the Galactic conversion factor  we 
obtained only for one Galaxy (NGC 6764) a molecular gas mass
larger than the atomic gas mass.
It is difficult to
assess the possible error caused by neglecting
molecular gas, however the above considerations have shown that
the likely error caused is less than a factor of 2. 

We estimate the total error in the dust-to-gas ratio
derived in this way to be about
a factor of 4. This number  takes into account the possible error in the dust
mass due to VSGs (factor 2), cold dust (factor 2, estimated from
the dispersion of the correlation in Fig. \ref{tau-mass}),
molecular gas (factor 2), and a varying \hi/optical diameter (factor 1.6).

For the dIrr sample there exists a good correlation 
between the dust-to-gas ratio and the metallicity (Fig. \ref{dust-gas-dirr}). 
Only one galaxy (GR8) is clearly off the correlation. 
Not taking into
account this galaxy and the  4 galaxies  with only upper limits 
in the FIR,   the correlation coefficient is $r=0.83$. 
The correlation is not linear;
a least square fit yields: 12+log(O/H) $\propto
(0.52\pm 0.25) \times\log(\mdust/\mhi)$. The error in the slope takes into
account the dispersion of the data points around the least-square fit
line, the
error in the dust mass of a factor 4 and an error in the metal abundance
of 0.2 dex. 
This result is in agreement with earlier studies:
Schmidt \& Boller (1993) obtained a very similar  slope
in the correlation as we (0.63 $\pm 0.25$) which is to be expected
because our sample is an extension (7 more galaxies) of theirs.
The results of Issa  \etal (1990) who compared the dust-to-gas
ratio in nearby galaxies are similar. From their
data a slope of 0.85 $\pm$ 0.1 can be obtained (Schmidt \& Boller 1993).

The sample of the BCDs shows no obvious correlation between the 
dust-to-gas ratio and the metallicity (Fig. \ref{dust-gas-bcd}). 
Both the average dust-to gas ratio as well
as the average metallicity tend to be higher for this sample than
for the  dIrrs (Fig. \ref{dust-gas-tot}).
For a given metallicity the dust-to-gas ratio is on average higher
than for the dIrr galaxies. 

\section{Model}

We are now interested in developing a simple model that explains
the observed behavior of the dust-to-gas/metallicity relation in dwarf galaxies.
In particular one would like to understand (i) the physics that governs
the observed relationship (ii) the apparently different trend
found in dIrrs and BCDs (iii) learn about star formation, metal production, and
mass loss history of these objects.

\subsection{Basic Equations}

To answer the above questions, we start writing the equations describing  
the rate of change of the galactic gas mass, $M_g$, the mass fraction of a given heavy
element $i$, denoted by $X_i=M_i/M_g$ ($M_i$ is the mass of the element $i$), 
and the ratio between the mass in element $i$ and the dust mass,
$f=M_d/M_i=M_d/X_iM_g$:
 
\begin{equation}
\label{mg}
{d\over dt}M_g= -\psi + E - W,
\end{equation}
\begin{equation}
\label{mi}
{d\over dt}M_i={d\over dt}[X_iM_g]= -X_i\psi + E_i  -X_i W,
\end{equation}
\begin{equation}
\label{md}
{d\over dt}M_d={d\over dt}[fX_iM_g]= -\alpha fX_i\psi + {f_{in}E_i }
 - {fX_i M_g\over t_{sn}} - \delta fX_i W.
\end{equation}
We now discuss in detail the various quantities entering 
eqs. \ref{mg}-\ref{md}.
The star formation rate (SFR) is denoted by $\psi$; $E$ is the total injection
rate from stars of all masses and ages; $E_i$ is the total injection rate of
element $i$ from stars; $W$ is the net outflow rate from the galaxy. The three
parameters in eq. \ref{md} $\alpha$, $f_{in}$ and $\delta$ refer to dust
properties: $\alpha f$ is the fraction of dust destroyed during star formation
($\alpha=1$ corresponds to destruction of only the dust incorporated into the
star, $\alpha > 1$ [$\alpha < 1$] corresponds to a net destruction [formation]
in the protostellar environment); $f_{in}$ is the value of the dust fraction in
the injected material; $\delta$ accounts for a possible different dust content
of the outflow with respect to the general interstellar medium ($\delta = 0
\Rightarrow$ no dust in the outflow; $\delta = 1 \Rightarrow$ the outflow is
as dusty as the ISM).
It is easy to understand the physical meaning of the above equations from
an inspection of the various terms. Eq. \ref{mg} and \ref{mi} describe the
detailed balance of the gas and $i$-element mass as due to star formation,
stellar injection and outflows; eq. \ref{md} accounts for the fact that dust
is destroyed/formed during star formation processes, it is injected by stellar
sources, destroyed by supernova shocks (on a characteristic 
time-scale $t_{sn}$) 
and lost in outflows. We have neglected the change of $M_i$ deriving from 
grain destruction in comparison with the stellar metal production;
this approximation holds for values ${\cal D}\simlt 10^{-3}$ and therefore
should be appropriate for our sample;
also eq. \ref{md}
does not take into account the possible formation or accretion of dust directly
in the 
interstellar medium. The validity of this assumption is difficult to assess,
since the determination of the grain accretion rate in the ISM would require
a detailed knowledge of the grain microphysics. However, since accretion is
well known to occur preferentially in high gas density regions (molecular
clouds), we hope that this process, due to the paucity of molecular gas  in
dwarfs galaxies, will not have a dramatic impact on our results. 

We next adopt the {\it Instantaneous Recycling Approximation}, (IRA) \ie stars
less massive than 1 $M_\odot$ live forever and the others die instantaneously, 
(Tinsley 1980)
which allows $E$ and $E_i$ to be written respectively as
\begin{equation}
E=R\psi;~~~~~~~~~~~~~~~~~~~~~~~  E_i=[RX_i + y(1-R)]\psi,
\end{equation}
(Note that Tinsley's eq. 3.14 contains a wrong extra factor $1-X_i$,
see Maeder 1992), where $R$ and $y$ are the standard returned fraction 
(the fraction of the total mass that has formed stars which is 
subsequently going to be ejected into the ISM)
and  net stellar yield (the total mass of an element $i$ ejected 
by all stars back into the ISM per unit mass of matter locked into stars), respectively.
The limitations of IRA are discussed by Tinsley (1980). Some contribution
to the dust mass can come from evolved stars, whose evolution
is only roughly described by the IRA approximation. However, since we will
show below that the major constituent of grains is oxygen -- almost totally 
contributed by massive stars, we estimate that IRA should suffices for our
purposes, even though the complete validity of this assumption should be
assessed by future studies.  
Moreover, we assume that the outflow rate is
$W = w\psi$, with $w={\rm const.}$, a common {\it ansatz} in chemical evolution models.  It is
also useful for our purpose to write the above equations in terms of the
dust-to-gas ratio, ${\cal D}=f \,X_i$. 
With these positions, eqs. \ref{mg}-\ref{md}
become  

\begin{equation}
\label{mg1}
{1\over \psi}{d\over dt}M_g= (R-1-w)
\end{equation}
\begin{equation}
\label{mi1}
{M_g\over \psi}{d\over dt}X_i = y(1-R)
\end{equation}
\begin{equation}
\label{md1}
{M_g\over \psi}{d\over dt}{\cal D}= -[R-1-w(1-\delta)+\alpha]{\cal D} +
f_{in}[RX_i + y(1-R)] - {M_g\over t_{sn}}{{\cal D}\over \psi}
\end{equation}

To determine $t_{sn}$ we follow McKee (1989), who writes this quantity as
\begin{equation}
\label{tsn}
t_{sn}={M_g \gamma^{-1}\over \bar\epsilon M_s(100)}
\end{equation}
where $\gamma$ is an ``effective'' supernova rate calculated including both 
Type II and Type I supernovae and accounting for the fact that not all the
SNRs interact with the galactic ISM. For sake of simplicity, and also because
the second effect should be less important in small objects like dwarf galaxies,
we neglect these complications and take $\gamma$ as the Type II SN rate.
We write $\gamma=\nu \psi$, where $\nu$ is the number of supernovae per unit
stellar mass formed. We assume $\nu$ to be the same as in the Galaxy, for which
$\psi \sim 3 M_\odot$~yr$^{-1}$ (Larson 1996) and $\gamma\sim0.022$~yr$^{-1}$ (van
den Bergh 1983); this yields $\nu = 1/136 M_\odot = 3.6 10^{-36}$~g$^{-1}$.
The mean efficiency of grain destruction by a shock is denoted by
$\bar\epsilon$; its precise value depends on the mean gas density and magnetic
field strength in the ambient medium. Lacking a precise determination of these
quantities in dwarf galaxies, we adopt the conservative value $\bar\epsilon=0.1$,
suggested by McKee (1989) for the Galaxy. Next we need an estimate for $M_s(100)$,
the gas mass shocked to a velocity of at least $100$~km~s$^{-1}$. From the
Sedov-Taylor solution in an homogeneous medium one can easily obtain:
\begin{equation}
\label{ms1}
M_s(100)=6800 \left({E\over 10^{51}}\right)\left({v_s\over 100 {\rm km~s}^{-1}}
\right)^{-2} M_\odot
\end{equation}
where $E$ is the energy release by the explosion and $v_s$ is the shock velocity.
We adopt the fiducial value 100 {\rm km~s}$^{-1}$ in eq. \ref{ms1} according to
McKee (1989). In conclusion, we have 
\begin{equation}
\label{tsn1}
t_{sn}={M_g \over \beta \psi}
\end{equation}
with $\beta=\bar\epsilon \nu M_s(100) \sim 5$. 

Since we are interested in the relation between $X_i$ and ${\cal D}$ in order
to compare it with the data, we eliminate time from the previous equations.
This yields 
\begin{equation}
\label{master}
{d{\cal D}\over dX_i}+a{\cal D} - bX_i = c,
\end{equation}
where
\begin{equation}
\label{coeffs}
a = {[R-1-w(1-\delta)+\alpha + \beta]\over y(1-R)},\qquad
b = {f_{in}R\over y(1-R)},\qquad
c=f_{in}.\qquad
\end{equation}
The solution of eq. \ref{master} is
\begin{equation}
\label{sol}
{\cal D}(X_i)={b\over a} X_i + (1-e^{-aX_i})({c\over a}-{b\over a^2})
\end{equation}
In order to quantify the above relation, we need to fix the value of 
$R$ and $y$ for the traced heavy element. We choose oxygen for the following
reasons: (i) it is produced mainly in Type II SNe, which are also the ones
responsible for the grain shock destruction; 
(ii) oxygen is  the main
constituent of grains (the relative abundances of atoms of O:C:Si:Mg in grains
are 100:54:6.5:5.2); (iii) a large sample of dwarf galaxies
with good quality abundance data for this species is available (see
Sect. 2). We therefore take $X_i\equiv X_o$, the oxygen mass fraction. 
The returned fraction and the net yield can be obtained using the 
standard formulae (Maeder 1992):
\begin{equation}
\label{yields}
R = {\int_{m_l}^{m_u} (m - w_m) \phi (m) dm \over
 \int_{m_l}^{m_u} m \phi (m) dm};\qquad  y = {1\over (1-R)}{\int_{m_l}^{m_u} 
m p_m \phi (m) dm \over
 \int_{m_l}^{m_u} m \phi (m) dm}
\end{equation}
where $\phi(m)$ is the initial mass function (IMF) 
defined between the lower and upper masses $m_l=1
M_\odot$ 
and $m_u=120 M_\odot$, and $w_m$ ($w_m=0.7 M_\odot$ for $m\le 4 M_\odot$ and
$w_m=1.4 M_\odot$ for $m > 4 M_\odot$) is the remnant mass (white dwarf or
neutron star).
We have used a power law form of the IMF, 
$m\phi(m)\propto m^{-x}$ with $x=1.35$ (standard Salpeter) and $x=1.7$ (Scalo 1986). 
We have taken the oxygen stellar yield, $p_m$ (\ie the mass fraction of a star of mass
$m$ converted into oxygen and ejected) from Arnett (1990). With these
assumptions we obtain $R=(0.79, 0.684)$ and $y=(0.0871, 0.0305)$ for the 
(Salpeter, Scalo) IMF, respectively.

As a consistency check of our simple chemical and dust evolution model
one could try to reproduce, at least qualitatively, the observed
($12+\log(O/H)-\log \mu$) relation, where $\mu=M_g/M_*$ is the ratio between the
gas and stellar mass of the system.  From eqs.\ref{mg1}-\ref{mi1}, it is easy
to obtain that
\begin{equation}
X_o = {y(1-R)\over (1-R-w)} \ln \left[{1+\mu+(w/(1-R))\over \mu}\right].
\end{equation}
The experimental determination of $\mu$ is very uncertain because 
it is difficult
to derive reliable values for $M_*$. However, we have been able to reasonably constrain
the value of $\mu$ for seven dIrrs in our sample 
from their H-band (1.6 $\mu$m) luminosity and we find $0.05 < \mu< 2$.
Using $w\sim 20-40$, the above formula reproduces quite satisfactorily the
observed trend and numerical value of the metallicity-astration relation.
This outflow rate, as we show in the next Section, is very close to the one
we require to explain the dust-to-gas ratio of dwarf galaxies.

\subsection{Model Results}

\begin{figure}
\psfig{file=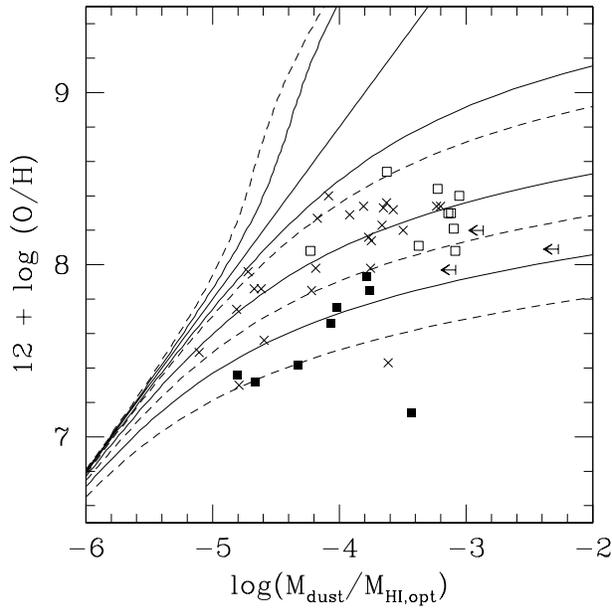,width=8.5cm,clip=,angle=0}
\caption{\label{model1} Comparison between the theoretical ${\cal D}-X_o$ curves
given by eq. \ref{sol} and data points for the entire sample
(symbols as in Fig. \ref{bol-temp}
and  \ref{dust-gas-dirr}).
Solid curves refer
to Salpeter IMF, dashed curves refer to Scalo IMF. Each set curves is
plotted for $f_{in}=0.01$ and the following values of 
$\chi=w(1-\delta)=0,5,10,30,90$, from top to bottom. }
\end{figure}

\begin{figure}
\psfig{file=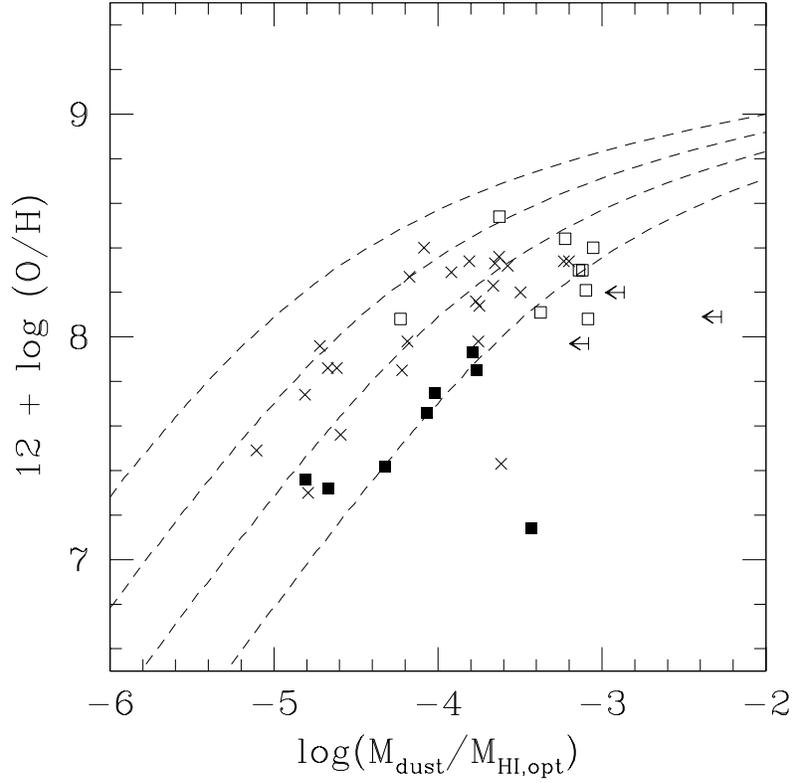,width=11.cm,clip=,angle=0}
\caption{\label{model2} Comparison between the theoretical ${\cal D}-X_o$ curves
given by eq. \ref{sol} and data points for the entire sample 
(symbols as in Fig. \ref{bol-temp}
and  \ref{dust-gas-dirr}) for different
values of $f_{in}=0.1, 0.03, 0.01, 0.003$, from bottom to top, 
a Scalo IMF and a fixed value of 
$\chi=10$.} 
\end{figure}

Once $R$ and $y$ have been fixed, the solution given by eq. \ref{sol} contains 
four free parameters: $\alpha, f_{in}, w, \delta$. However, the solution is
quite insensitive to $\alpha$: we take it equal to unity.
Thus, we are left with  three free parameters.
The precise value of  
$f_{in}$, the dust-to-O mass ratio in stellar dust sources, 
is rather uncertain; a simple estimate can be obtained as follows.
A typical SN ejects $\sim 4 M_\odot$ of heavy elements into the ISM. However, only 
the fraction of oxygen (which dominates the ejected material) that can 
bind to refractory elements can go into grains,
at most 25\% as derived from ISM depletion studies. In addition a rather
optimistic upper limit for the efficiency of grain formation (largely unknown) 
is $\sim 40\%$. Thus, $f_{in} \le 0.1$, but likely to be typically smaller than
this value.
Given our present ignorance of the processes governing dust formation,
we will explore below (see Fig. \ref{model2}) the effects of varying 
$f_{in}$ within a range of reasonable values.
%
%
For the last two free parameters,
which describe the outflow process, \ie
$w$ and $\delta$, the outflow rate (in units of the SFR) and the dust fractional
content of the outflow,
we take $0 < w < 90$ 
and $0 < \delta < 1$ in order to cover a reasonably large parameter space.
Fig. \ref{model1} shows the main results of this Section.
 There we plot the solutions
for different values of $w$ and $\delta$  and compare them with the data 
points corresponding to the dust-to-gas vs. metallicity relation  shown in Fig. \ref{dust-gas-tot}. The model results are shown for both a 
Salpeter (solid lines) and a Scalo (dashed) 
IMF.
Several interesting points can be made from an inspection of this Fig. 
\ref{model1}.

First, we conclude that
a simple power-law does not represent a good fit to the data, 
but a more much physically meaningful functional is the
one given by eq. \ref{sol}.
For low values of ${\cal D}$ the relation is shown to be quasi-linear,
whereas for higher values the shape of the curve depends sensitively
on $\delta$ and $w$, through the coefficient $a$ in eq. \ref{sol}.
In general, for a fixed value of $\delta$, a higher value of the
outflow rate produces a flatter curve; the same effect can be obtained by 
decreasing $\delta$: both situations allow the system to retain a larger
fraction of dust. However, the solutions only depend on  
the product $\chi=w(1-\delta)$, as can be seen from the coefficient 
$a$ in eq. \ref{coeffs}; $\chi$ can be interpreted as a ``differential''
outflow rate.

Next, we note that the ${\cal D}-X_o$ relation {\it does not} depend on the 
particular SFR $\psi$ of the galaxy: this could be expected since both metal
and dust production are strictly connected implying that the relation found
is rather general; however, if all the parameters are constant in time,
a given galaxy
is forced to move along the appropriate curve as it gets metal and dust
enriched. 
The curves are moderately sensitive to the IMF, with  the Scalo IMF covering
a larger region of the ${\cal D}-X_o$ plane for the given set of parameters. 

The results also point out the importance of outflows for dwarf galaxies. 
The effect of mass loss is particularly evident in the high-${\cal D}$ 
part of the curves where it results in a flattening;
the data require that most of the dwarfs in the sample
must have undergone a substantial outflow episode. 
The segregation of BCDs in the high metallicity,
high dust-to-gas ratio region of the plane can be then explained by a higher
SFR with respect to dIrrs, which implies a more substantial enrichment in 
metals and dust. Support 
to this hypothesis is lent by the fact that BCDs are all 
consistent with large values ($>20$)
of the differential outflow rate $\chi$, since ultimately the
SFR determines the rate of SNe powering the wind. 

Some uncertainty affects 
the low-${\cal D}$, low-$X_o$ part of the curves which is sensitive to the
precise value of $f_{in}$, as shown by Fig. \ref{model2} for the Scalo IMF. 
In the limit $X_i\rightarrow 0$, eq. \ref{sol} reduces to ${\cal D}(X_i)
\simeq cX_i = f_{in}X_i$, \ie the solution scales linearly
with $f_{in}$, which we have allowed to vary in $0.003 \le f_{in} \le 0.1$.   
Thus the precise value of $f_{in}$ could be possibly constrained by 
observations of extremely metal poor objects.
 

\section{Summary and Conclusions}

We have compared the metallicity and the dust-to-gas ratio in a sample of 44 
(28 dIrr + 16 BCD) dwarf  galaxies.
The dust mass has been derived from the FIR emission assuming
that the dust composition is the same for all galaxies. We have
carried out various tests in order to check the reliability  of
this dust mass determination.
For the sample of dIrr galaxies we found a good correlation
between the two quantities, obeying the empirical relation  12+log(O/H)
$\propto (0.52\pm 0.25) \times
\log(\mdust/\mhi)$.
 For BCD galaxies we did not find such a good correlation; 
both the dust-to-gas ratio and the metallicity tend
to be higher than in the dIrr sample, and       
for a given metallicity the dust-to-gas ratio is on average higher
than for the dIrr galaxies. 
 
We have derived a simple analytical form for the expected dust-to-gas ratio vs.
metallicity, ${\cal D}-X_o$,  relation (we have assumed that oxygen is a good
metallicity indicator) and compared it with the observational data. 
The main results are the following:

$\bullet$ For low values of ${\cal D}$ the ${\cal D}-X_o$ 
relation is  shown to be quasi-linear, whereas for higher values the curve 
strongly deviates from the linear behavior. 

$\bullet$ The deviation from the linear behavior depends critically on 
the 
parameter $\chi=w(1-\delta)$, the ``differential'' mass outflow rate from 
the galaxy in units of the SFR $\psi$,  
and is negative if $\chi < 5$. Most of the galaxies in the sample can then 
be inferred to have undergone a substantial outflow episode.

$\bullet$ The {\it shape} of the ${\cal D}-X_o$ curve does not depend on the 
SFR, $\psi$, of the galaxy but only on the outflow rate; however, the 
specific location of a given galaxy on the curve does depend on $\psi$.

$\bullet$ The BCD metallicity segregation is due to a higher SFR, also suggested
by their consistency with the generally larger 
values of the outflow rates required.
The lack of correlation could then be produced by largely different 
star formation rates and values of $\chi$ in these objects.

\acknowledgements{
We would like to thank E. Dwek and C. Xu for
helpful and interesting discussions and an anonymous referee
for useful suggestions. U. L. gratefully
acknowledges the receipt of a grant by the Deutsche 
Forschungsgemeinschaft (DFG) and by the Comisi\'on Interministerial de
Ciencia y Technolog\'\i a (Spain). This research has made use
of the NASA/IPAC extragalactic database (NED) and of the 
Lyon-Meudon Extragalactic Database (LEDA).
}

\vfill\eject

%
\begin{table}
\caption{Data of the dwarf irregular sample}
\begin{tabular}{lccccccc}
\tableline
\tableline
Name & D & $\fsix$ & $\fhun$ &  $\log(\mdust$) & $\log(\mhiopt)^{(1)}$ &
$\log({\mdust\over\mhiopt})^{(1})$ & $12+\log(O/H)$ \\  
  & [Mpc]  & [Jy] & [Jy] &  [$\msun$]  & [$\msun$] &
    & \\
\tableline
Gr 8    &  2.20&     0.020&     0.143& 3.06& 6.68& -3.61&7.43 \\ 
Leo A   &  1.60&     0.090&     0.270& 2.50& 7.29& -4.79&7.30 \\
WLM     &  1.00&     0.320&     1.040& 2.73& 7.54& -4.81&7.74 \\ 
SMC     &  0.06&  6688.910& 15021.930& 4.20& 8.39& -4.19&7.98 \\ 
LMC     &  0.05& 82917.000&184686.690& 5.13& 8.33& -3.20&8.34 \\ 
DDO 47  &  2.00&     0.125&     0.555& 3.26& 7.48& -4.22&7.85 \\ 
Sex A   &  1.50&     0.503&     0.849& 2.56& 7.67& -5.11&7.49 \\ 
Sex B   &  1.70&     0.246&     0.684& 2.90& 7.50& -4.59&7.56 \\ 
IC 10   &  2.60&    31.230&    71.250& 5.16& 8.66& -3.50&8.20 \\ 
IC1613  &  0.70&     1.420&     3.690& 2.82& 7.44& -4.62&7.86 \\ 
IC2574  &  2.40&     2.410&    10.620& 4.70& 8.45& -3.75&7.98 \\ 
IC4662  &  2.20&     7.360&    11.230& 3.95& 7.87& -3.92&8.29 \\ 
IC5152  &  3.00&     2.461&     6.861& 4.40& 8.03& -3.63&8.36 \\ 
NGC55   &  2.00&    77.000&   174.100& 5.32& 8.97& -3.65&8.33 \\ 
NGC1156 &  6.50&     5.170&     9.200& 4.91& 8.57& -3.66&8.23 \\ 
NGC1569 &  2.20&    39.223&    31.748& 3.98& 7.75& -3.77&8.16 \\ 
NGC2366 &  3.30&     3.303&     4.578& 3.85& 8.57& -4.72&7.96 \\ 
NGC3510 &  8.90&     0.598&     1.457& 4.58& 8.67& -4.08&8.40 \\ 
NGC4214 &  4.10&    14.470&    25.470& 4.94& 8.75& -3.81&8.34 \\ 
NGC4236 &  4.90&     3.980&    10.020& 4.92& 9.10& -4.17&8.27 \\ 
NGC4449 &  3.90&    14.762&    37.806& 5.31& 8.89& -3.57&8.32 \\ 
NGC5253 &  6.90&    30.510&    29.360& 5.06& 8.29& -3.23&8.34 \\ 
NGC5408 &  4.20&     2.360&     2.160& 3.46& 8.13& -4.67&7.86 \\ 
NGC6822 &  0.50&    47.630&    95.420& 3.77& 7.52& -3.75&8.14 \\ 
SDIG    &  2.90&    $<$ 0.093&  $<$ 0.159& 2.42& 6.74& -4.32&7.42 \\ 
UGC4483 &  4.00&    $<$ 0.079&  $<$ 0.149& 2.73& 7.40& -4.67&7.32 \\ 
DDO 167 &  3.50&    $<$ 0.093&  $<$ 0.183& 2.73& 6.80& -4.07&7.66 \\ 
DDO 187 &  2.60&    $<$ 0.084&  $<$ 0.134& 2.20& 7.01& -4.81&7.36 \\ 
\tableline
\end{tabular}

$^{(1)}$ $\mhiopt=\mhi/2$ refers to the \hi mass inside the optical
diameter (see text)
\end{table}
\begin{table}
\caption{Data of the Blue Compact Dwarf sample}
\begin{tabular}{lccccccc}
\tableline
\tableline
Name & D & $\fsix$ & $\fhun$ &  $\log(\mdust$) & $\log(\mhiopt)^{(1)}$ &
$\log({\mdust\over\mhiopt})^{(1)}$ & $12+\log(O/H)$ \\
  & [Mpc]  & [Jy] & [Jy] & [$\msun$] & [$\msun$] 
 &  &  \\
\hline
UM 448  &  6.00&     4.140&     4.321& 4.16& 7.24& -3.09&8.08 \\
IC 3258 & 21.20&     0.490&     0.970& 5.03& 8.25& -3.22&8.44 \\ 
Mrk 7   & 42.30&     0.480&     0.970& 5.64& 9.26& -3.63&8.54 \\ 
Mrk 33  & 21.60&     4.680&     5.300& 5.41& 8.47& -3.05&8.40 \\ 
Mrk 35  & 14.50&     4.950&     6.740& 5.29& 8.43& -3.14&8.30 \\ 
Mrk 450 & 12.10&     0.480&     0.820& 4.37& 7.47& -3.10&8.21 \\ 
NGC4670 & 12.10&     2.630&     4.470& 5.10& 8.22& -3.12&8.30 \\ 
NGC4861 & 12.90&     1.970&     2.260& 4.60& 8.83& -4.23&8.08 \\ 
II Zw70 & 17.60&     0.710&     1.240& 4.89& 8.26& -3.37&8.11 \\ 
UM 455  & 51.10& --$^{(2)}$  & --$^{(2)}$ & 5.59& 9.61& -4.02&7.75 \\ 
Mrk 600 & 14.00& --$^{(2)}$  & --$^{(2)}$& 4.46& 8.23& -3.76&7.85 \\
IZw 18  & 11.00& --$^{(2)}$  &  --$^{(2)}$ & 4.25& 7.68& -3.43&7.14 \\ 
IZw 36  &  4.80& --$^{(2)}$  &  --$^{(2)}$& 3.53& 7.32& -3.79&7.93 \\ 
IIZW40  & 10.10&     6.020&    $<$19.700& 6.02& 8.29& -2.27&8.09 \\ 
Mrk750  & 12.40&     0.440&    $<$0.840& 4.47& 7.34& -2.86&8.20 \\ 
IZw123  & 10.40&     0.300&    $<$0.640& 4.28& 7.36& -3.08&7.97 \\ 
\tableline
\end{tabular}
$^{(1)}$ $\mhiopt=\mhi/2$ refers to the \hi mass inside the optical
diameter (see text)

$^{(2)}$ not detected by IRAS 
\end{table}

\begin{references}


Andreani, P., Casoli, F. \& Gerin, M. 1995, A\&A, 300, 43

Arnett, D. W. 1990, in Chemical and Dynamical Evolution of Galaxies, eds. F.
Ferrini \etal (Pisa: ETS), 409


Borkowski, K. J. \& Dwek. E. 1995 ApJ, 454 254

Bouchet, P., Lequeux, J., Maurice, E., Pr\'evot, L.\& 
Pr\'evot-Burnichon, M.L. 1985, A\&A 149, 330 

Campbell, A., Terlevich, R.\& Melnick, J. 1986, MNRAS, 223, 811

Clayton, G.C., Green, J., Wolff, M.J., Zellner, N.E.B., Code, A.D., 
Davidsen, A.F., \etal 1996, ApJ, 460, 313

Cohen, R.S., Dame, T.M., Garay, G., Montani, J., Rubio, M.\& Thaddeus, P.
1988, ApJ, 331, L95

Deharveng, J. M., Sasseen, T. P., Buat, V., Bowyer, S., Lampton, M. \& Wu, X.
1994, A\&A, 289, 715

D\'esert, F. X., Boulanger, F. \& Puget, J. L. 1990, A\&A, 237, 215

de Vaucouleur, G., de Vaucouleur, A., Corwin H., \etal, 1991,
Third Reference Catalog of Galaxies, (New York: Springer) (RC3)

Devereux, N. A. \& Young, J. S. 1990, ApJ, 359, 42 

Donas, J. \& Deharveng, J. M. 1984, A\&A, 140, 325

Donas, J., Deharveng, J. M., Laget, M., Milliard, B. \& Huguenin, D. 1987,
A\&A, 180, 12


Dwek, E. \& Scalo, J. M. 1980, ApJ, 239, 193

Elfhag, T., Booth, R. S., H\"oglund, B., Johansson, L. E. B.\& 
Sandqvist, Aa. 1996, A\&AS, 115, 439

Fanelli, M. N., Waller, W. W., Smith, D. A., Freedman, W. L., 
Madore, B., Neff, S. G., O'Connell, R. W., Roberts, M. S., Bohlin, R.,
Smith, A.M. \& Stecher, T. P. 1997, ApJ, 481, 735 

Ferrara, A., Ferrini, F., Franco, J., \&  Barsella, B. 1991, ApJ, 381, 137

Fitzpatrick, E.L. 1986, AJ, 92, 1068

Fouqu\'e, P. 1983, A\&A, 122, 273

Franco, J. \& Cox, D. P. 1986, PASP, 98, 1076


Hildebrand, R. H. 1983, QJRAS, 24, 267

Hill, R. S., Home, A. T., Smith, A. M., Bruhweiler, F. C.,
Cheng, K. P., Hintzen, P. M. N. \& Olivsersen, R. J. 1994, ApJ, 430, 568

Huchtmeier, W. K. \& Richter O.-G. 1989, A General Catalogue of HI 
Observations of Galaxies, (New York: Springer)

Huchtmeier, W.K., Hopp, U. \& Kuhn, B. 1997, A\&A, 319, 67

Hunter, D. A., Gallagher, J. S. \& Rautenkranz, D. 1982, ApJS, 49, 53

Hunter, D. A., Gallagher, J. S., Rice W. L., \& Gillett, F. C. 1989,
ApJ, 336, 152

Hunter, D. A., Hawley, W. N. \& Gallagher, J. S. 1993, AJ, 106, 1797

Jones, A. P., Tielens, A. G. G. M., Hollenbach, D. J. \& McKee, C. F. 1994,
ApJ, 433, 797

IRAS Faint Source Catalog, 1990, Version 2.0,
Moshir M. \etal Infrared Processing and Analysis Center

Issa, M. R., MacLaren, I. \& Wolfendale, A. W. 1990, A\&A, 236, 237

Kennicutt, R. C. \& Kent, S. M. 1983, AJ, 88, 1094

Koornneef, J., 1982 A\&A, 107, 247

Kozasa, T, Hasegawa, H. \& Nomoto, K. 1989, ApJ, 344, 325


Kunth, D. \& S\`evre, F. 1986, in  Star forming dwarf galaxies 
and related objects, eds. Kunth, Thuan \& Tran-Thanh, 
(Gif-sur-Yvette: Fronti\`ere)

Larson, R. B. 1996, in The Interplay between Massive Star Formation, the ISM and
Galaxy Evolution, 11th IAP Meeting, eds. D. Kunth \etal (Gif-sur-Yvette:
Fronti\`ere), 3

Lisenfeld, U., V\"olk, H.J. \& Xu, C. 1996, A\&A, 306, 677

Maeder, A. 1992, A\&A, 265, 105


Masegosa, J., Moles, M. \& Campos-Aguilar, A. 1994, ApJ, 420, 576 

McGaugh, S. 1991, ApJ, 380, 140

McKee, C. F. 1989, in Interstellar Dust, IAU Symp. 135, eds. L. J.
Allamandola \& A. G. G. M. Tielens, (Dordrecht: Kluwer), 431 

Melisse, J. P. M. \& Israel, F. P. 1994a, A\&AS, 103, 391

Melisse, J. P. M. \& Israel, F. P. 1994b, A\&A, 285, 51

Moseley,  S. H., Dwek, E., Glaccum, W., Graham, J. R., Loewenstein, R. F. \&
Silverberg, R. F. 1989, Nature, 340, 697

Norman, C. A. \& Ikeuchi, S. 1989, ApJ, 345, 372


Rowan-Robinson, M. 1992, MNRAS, 258, 787

Rubio, M., Garay, G., Montani, J.\& Thaddeus, P. 1991, ApJ, 368, 173

Scalo, J. M. 1986, Fund. Cosm. Phys., 11, 1

Schmidt K. H. \& Boller T. 1993, Astron. Nachr., 314, 361

Skillman, E. D., Kennicutt, R.C. \& Hodge, P. W. 1989, ApJ, 347, 875

Sodroski, T.J., Dwek, E., Hauser, M. G. \& Kerr, F.J. 1987, ApJ, 322, 101

Staveley-Smith, L., Davies, R.D., Kinman, T.D., 1992, MNRAS, 258. 334

Tervelich, R., Melnick, J., Masegosa, J. \& Moles, M. 1992, A\&AS, 91, 285

Tinsley, B. M. 1980, Fund. Cosm. Phys., 5, 287

van den Bergh, S. 1983, PASP, 95, 388


Xu, C. \& de Zotti G. 1989, A\&A, 225, 12

Xu, C.\& Buat, V. 1995, A\&A, 293, L65

Xu, C.\& Helou, G. 1996, ApJ, 456, 163


Young, J. S., Xie, S., Kenney, J. D. \& Rice W. L. 1989, ApJS, 261. 492

Young, J. S, Shuding, X., Tacconi, L., \etal 1995, ApJS, 98, 219
\end{references}
\end{document}